\documentclass{Interspeech2024}

\interspeechcameraready

\title{Continuous Learning of Transformer-based Audio Deepfake Detection}

\name[affiliation={1,2}]{Tuan Duy}{Nguyen Le}
\name[affiliation={2}]{Kah Kuan}{Teh}
\name[affiliation={2}]{Huy Dat}{Tran}

\address{
  $^1$Viettel AI, Viettel Group\\
  $^2$Aural and Language Intelligence Department, Institute for Infocomm Research, A*STAR, Singapore 
  }
\email{tuanduynl129@gmail.com, tehkk@i2r.a-star.edu.sg, hdtran@i2r.a-star.edu.sg}
\usepackage{amsmath,graphicx,multirow,booktabs,tabularx,caption}


\begin{document}

\maketitle

\begin{abstract}
    
This paper proposes a novel framework for audio deepfake detection with two main objectives: i) attaining the highest possible accuracy on available fake data, and ii) effectively performing continuous learning on new fake data in a few-shot learning manner. Specifically,  we conduct a large audio deepfake collection using various deep audio generation methods. The data is further enhanced with additional augmentation methods to increase variations amidst compressions, far-field recordings, noise, and other distortions. We then adopt the Audio Spectrogram Transformer for the audio deepfake detection model. Accordingly, the proposed method achieves promising performance on various benchmark datasets. Furthermore, we present a continuous learning plugin module to update the trained model most effectively with the fewest possible labeled data points of the new fake type. The proposed method outperforms the conventional direct fine-tuning approach with much fewer labeled data points.
\end{abstract}

\section{Introduction}
\label{sec:intro}

Audio deepfakes pose a significant and growing threat to humanity by creating fraudulent and misleading content that can deceive people or influence public opinion. \cite{review}. More recently, they have been utilized for cybersecurity attacks targeting commercial applications. \cite{spoofing-review}. These audio deepfakes can be generated using various AI techniques, including text-to-speech (TTS), voice transformation or cloning (VC), and, more recently, audio large language models (LLMs) for text-to-audio conversion.

\noindent Research in audio deepfake detection is currently concentrated on devising effective methods from a given limited training data. Public contests \cite{asv17}\cite{asv19}\cite{asv21} have been attracting research works from many universities and companies and subsequently set up a solid research community in audio deepfake detection. Baseline methods have been reported in the literature, from specific time-frequency representation \cite{wavelet} to specific neural network designs \cite{ASSERT} \cite{resnet34}. More recent works adopt large SSL pre-trained models such as wav2vec2.0 \cite{wav2vec2} or WavLM \cite{wavlm} for audio deepfake detections and achieved good results on challenges’ data \cite{wavlmdaf}. A shortcoming of challenges’ driven solutions is that it is not designed to be an industrial solution, trained on relatively small data sources, and hence not consistent enough in open test conditions. More recent initiatives like the Kaggle Deep Fake Detection Challenge\footnote{https://www.kaggle.com/competitions/deepfake-detection-challenge}, and AISG Trusted Media Challenge \cite{AISG} are more toward industrial solutions. However, these challenges have primarily focused on video or multimodal analysis, making them less suitable for assessing audio deepfake detection capabilities. Another significant problem of audio deepfake detection which has less been discussed in the challenges or public benchmarks is the fact that the original task is a specific open-set problem, with both classes (i.e. bonafide fake) open, but the current works are more on supervised scenarios. With the new fake models and techniques that have been developed frequently, a more challenging issue is how to design the solution to be able to update amid the presence of new fake types in the fastest manner using the fewest labeled data.

\noindent In this paper, we try to address two above scopes of an industrial solution for audio deepfake detection, which can be summarised as follows: 1) achieving the highest possible accuracy on available datasets; 2) being able to effectively perform continuous learning on new fake data with very few training samples. To construct this system, we first collect a huge amount of audio deepfake data from different TTS, VC, and audio LLM, reported in the research communities. Our data includes more than 2 million fake samples from more than 50 published speech-generation open sources. Then we employ several data augmentation techniques to increase the signal quality variations corresponding to compression, far-field, noises, and distortions. We adopt Audio Spectrogram Transformer (AST) deep learning architecture to build the model, which is initialized with big Audioset supervised audio classification task \cite{ast}. The proposed model has achieved state-of-the-art (SOTA) performance on almost all the public datasets. Next, we develop a novel continuous learning plugin, to address the model updating for a new fake type using the fewest labeled data. Unlike, conventional methods performing fine-tuning operations, our updating plugin includes two stages: in the beginning, we adopt discriminative learning based on trained AST model embedding with gradient boost machine \cite{gradient-boosting}, which is shown to be superior to fine-tuning in detecting new fake data with very small number of labeled data.  Once we have collected enough new fake samples from unsupervised sources, at the second stage, we apply the conventional fine-tuning of the operating deep learning model accordingly. The proposed plugin method has shown to be much more effective than conventional fine-tuning approaches. For unseen datasets, the proposed continuous learning system can improve the AUC from an initial 70+\% to 90+\%, with just 0.1\% of the training databases, and further elevate it to more than 95\% after performing unsupervised detection and model fine-tuning.

\noindent The organization of the paper is as follows: next, in Section \ref{sec2}, we will give an overview and experimental of our Transformer based system for detecting deepfake. Section \ref{sec3} then presents the continuous learning system to deal with new fake data. Finally, Section \ref{sec4} concludes the work.

\begin{figure}[t]
  \centering
  \includegraphics[width=\linewidth]{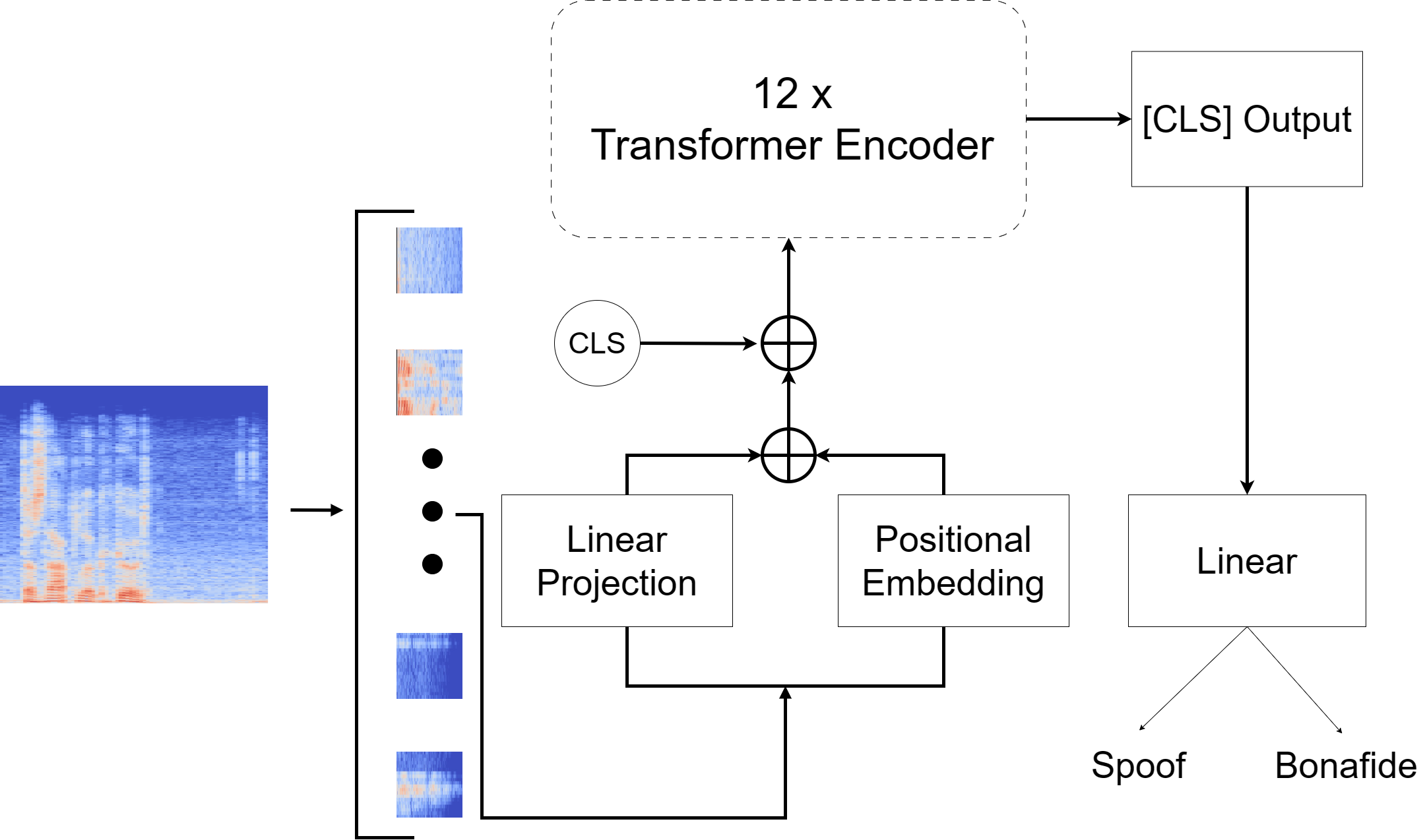}
  \caption{The architecture of the proposed method.}
  \label{fig:ast}
\end{figure}

\section{Transformer-based Deepfake Audio Detection}
\label{sec2}
\subsection{Motivation}
\noindent Building on the successes of visual intelligence, the most recent approach to classifying deepfake audio utilizes a CNN-based model to capture local spectral information from representations such as spectrograms or LFCC. Recently, methods working directly on raw audio waveforms have achieved state-of-the-art results on current anti-spoofing benchmark datasets. However, it remains uncertain whether these approaches are specifically addressing a particular dataset without proving their effectiveness for the broader audio deepfake problem. Furthermore, processing audio data in the raw waveform can be computationally costly; as a result; impractical.

\noindent With the emergence of Transformers \cite{self-attention}, which have achieved remarkable results in the natural language processing domain, exploring the adaptation of this architecture for audio appears promising. The success of Transformers largely depends on the foundation of pre-trained models, which necessitates massive datasets, making the application of this architecture to our problem challenging. Recently, the rise of large pre-trained foundation models using self-supervised learning (SSL) in the speech domain, such as wav2vec 2.0 \cite{wav2vec2}, WavLM \cite{wavlm}, and HuBERT \cite{hubert}, has garnered significant attention. While most of these works focus on improving ASR performance, several approaches have been proposed for fine-tuning these speech foundation models on downstream tasks, such as Speech Emotion Recognition or Speaker Verification \cite{finetunewav2vec2}. In addition, research has found that scaling supervised pre-training can significantly boost performance without the use of self-supervision techniques \cite{whisper}. Therefore, benefiting from the advantage of having collected a vast amount of data in the domain, this research explores the problem of audio deepfake detection through the following initiatives:
\begin{itemize}
    \item Proposing a large-scale trained Audio Spectrogram Transformer as a new state-of-the-art system for audio deepfake detection.
    \item Training a robust AST model that is effective on low-resolution audio qualities.
    \item Developing a plug-in module to perform continuous learning of new fake types.
\end{itemize}

\subsection{System Description}
Fig. \ref{fig:ast} illustrates our proposed system. Specifically, we adopted the AST, originally used for Audio Event Detection on the large-scale Audioset dataset. Inspired by Vision Transformer, AST addresses the challenge of applying Transformers, which are typically used for sequences of data, to a 2D audio spectrogram. This is achieved by splitting the spectrogram into sequences of $16 \times 16$ patches with a 6-pixel overlap, meaning a stride of 10 in both the time and frequency dimensions. In our system, we decided not to overlap patches in the spectrogram to avoid difficulties when dealing with samples containing both real and fake signals. For stability and robustness, we use a log mel filter bank with a $25ms$ Hamming window and a $10ms$ sliding length, with $128$ triangular mel-frequency bins. Consequently, we create a $128 \times 100t$ spectrogram, where $t$  denotes the audio length in seconds, thus the total patches will be $8 \times [100t/16] $. Each patch is then flattened and fed into the projection layer to obtain a 768-dimensional embedding. Generally, each input sample will consist of a $CLS$ token along with a sequence of patch embeddings and their positional embeddings to maintain order information before being fed into the Transformer Encoder block. The system comprises 12 Transformer Encoder blocks, taking the output of the $CLS$ token onwards to a fully connected layer to map the representation into labels, specifically here into bonafide and fake classes.

\subsection{Datasets}
\begin{table}[htbp]
\centering
\caption{Training data summary}
\label{traindata}
\begin{tabular}{|c|c|c|c|}
\hline
Type of samples    &Bonafide  & VC & TTS  \\
\hline\hline
Number of Speakers     &3357   &544     &1621    \\
\hline
Number of Systems      & -    &26     &27   \\
\hline
Number of Samples      &913849   &432865     &944363   \\
\hline
\end{tabular}
\end{table} 

\noindent We have collected large-scale training data including more than two million samples from multiple speech generation sources. Details of training data is shown in Table \ref{traindata}

\noindent One of the aims of the system is to provide robustness for low-audio-quality samples. Therefore, we apply different bandpass filters and lossy compression to the training samples for augmentation. Roughly 10\% of the total training samples are used for each low-resolution audio generation technique. Besides, we also use common data augmentation techniques, including mixup with a ratio of 0.5 and SpecAug with mask time and frequency dimensions of 192 frames and 48 bins, respectively.

\noindent In evaluation, we use three benchmark datasets that have not been used during the training.
\begin{itemize}
    \item The ASVspoof2019 LA evaluation dataset \cite{asv19} contains 7,355 real samples from 67 speakers and 63,882 audio deepfake samples from 48 speakers. There are totally 19 different audio deepfake generation methods, both seen and unseen techniques compared with its training subset.
    \item Secondly, FakeAvCeleb \cite{fakeavceleb} is a novel multimodal dataset that generated not only fake audio but also fake video. For this experiment, we only care about classifying between fake and real audio of each sample. The datasets contain 10209 real audio samples and 11335 fake ones, with the fake audio generated using SV2TTS from the real one.
    \item Finally, to test the model on a realistic unseen scheme, we use the in-the-wild dataset \cite{itw}, which contains 20.8 hours of bonafide and 17.2 hours of spoofed audio from 58 different speakers, who are celebrities and politicians.
\end{itemize}      
\subsection{Result Analysis}
\subsubsection{Result on standard dataset}
\noindent Table \ref{tab:itw} presents a comparative analysis of the performance between various baseline systems and our proposed approach on the ASVspoof 2019 evaluation, as well as on the In-the-Wild datasets, with baseline results sourced from \cite{itw}. 
\begin{table}[htbp]
  \caption{EER(\%) performance comparison of different systems on ASVspoof 2019 and In-the-wild dataset}
  \label{tab:itw}
  \centering
  \begin{tabular}{lll}
    \toprule
    \textbf{Model}      & \textbf{ASVspoof2019}    & \textbf{In-the-Wild}                  \\
    \midrule  
    LCNN &6.35  &65.56\\
    LCNN-LSTM &6.23  &61.5\\
    LCNN-Attention &6.76  &66.68\\
    MesoNet &7.42  &46.94\\
    MesoInception &10.02  &37.41\\
    ResNet18 &6.55  &49.76\\
    \midrule
    CRNNSpoof &15.66  &41.71\\
    RawPC  &3.09  &45.72\\
    RawNet2 &3.15  &33.94\\
    RawGAT-ST &\textbf{1.23}  &37.15\\
    \midrule
    Our AST &4.06   &\textbf{27.78}\\
    \bottomrule
  \end{tabular}
\end{table}

Table \ref{tab:fakeavceleb} highlights the performance of our system relative to other baselines on the FakeAVCeleb datasets, with findings drawn from \cite{fakeavceleb} \cite{deepfakeSV} \cite{fakeavcelebevaluation}, which represents the top current system that relies solely on audio to our knowledge. 
\begin{table}[htbp]
  \caption{Performance comparison of different systems on FakeAVCeleb dataset. The evaluation metrics are chosen based on previous works for better comparison}
  \label{tab:fakeavceleb}
  \centering
  \begin{tabular}{lll}
    \toprule
    \textbf{Model}      & \textbf{Accuracy(\%)}     & \textbf{AUC(\%)}           \\
    \midrule
    MesoInception-4                    & 54.0     & 73.5                                  \\
    Xception               & 76.3           & -                \\
    EfficientNet-B0                    & 50.0      & -                           \\
    VGG16          & 67.1             & 97.8        \\
    LCNN-LSTM          & -             & 67.3        \\
    RawNet2          & -             & 45.5        \\
    H/ASP          & -            & 94.5        \\
    ECAPA-TDNN          & -             & 90.4        \\
    POI-Forensics          & -             & 86.1        \\
    \midrule
    Our AST           & \textbf{92.6 }         &     \textbf{99.9}                 \\
    \bottomrule
  \end{tabular}
\end{table}
\noindent Our method achieved an Equal Error Rate (EER) of 4.06\% on ASVspoof 2019, surpassing all spectrogram-based methods and matching the performance of state-of-the-art (SOTA) systems utilizing raw waveform inputs. In experiments conducted in an open set setting, the results demonstrate that AST significantly surpasses all prior work on FakeAVCeleb and In-the-Wild datasets. This underscores the superior generalizability of AST, indicating its capability to detect certain fake methods even without having encountered similar samples during training. This effectiveness can be attributed to the observation that some Text-to-Speech (TTS)/Voice Conversion (VC) methods share common elements such as Mel-spectrograms and Vocoders, which the model may have learned to identify from other fake samples.

\subsubsection{Result on low-resolution dataset}
\noindent Synthetic speeches generated with malicious intentions are shared on social platforms such as YouTube, resulting in low-resolution audio qualities caused by various types of codecs. One of our goals is to build a robust solution that can detect audio deepfakes from both high- and low-resolution audio sources. Table  \ref{tab:fakeavceleb-low} reports experimental results for different types of audio low resolutions caused by built-in filtering and compression modules, for example, in MP3 format. 
\begin{table}[htbp]
\centering
    \caption{Effect of data augmentation on Original and Low-Resolution FakeAVCeleb dataset}
    \label{tab:fakeavceleb-low}
  \begin{tabularx}{\linewidth}{lcccc}
    \toprule
    \multirow{2}{*}{\textbf{Dataset}}      & \multicolumn{2}{c}{\textbf{Without Augmentation}}     &\multicolumn{2}{c}{\textbf{With Augmentation}} \\
    \cmidrule(lr){2-3}
    \cmidrule(lr){4-5}
                        & \textbf{Accuracy}     & \textbf{AUC} & \textbf{Accuracy}     & \textbf{AUC}          \\
    \midrule
    Original           & 71.8          &     93.8  & 92.6          &     99.9               \\
    Bandpass           & 60.6          &     91.2  & 89.1          &     96.2               \\
    Compress           & 48.8          &     63.9  & 87.3          &     93.7               \\
    \bottomrule
  \end{tabularx}
\end{table}
We observe that our model, trained with specifically designed augmentation techniques, greatly improves the detection accuracies, particularly for compression, from 63.9\% to 93.7\% AUC.

\section{Continuous Learning of Deepfake Audio Detection}
\label{sec3}
\begin{table*}[t]
\centering
    \caption{EER (\%) results on our continuous learning approach. Baseline indicates current performance before continuous learning}
    \label{tab:continuous}
  \begin{tabular}{cccccccc}
    \toprule
    \multirow{2}{*}{\textbf{Dataset}} & \multirow{2}{*}{\textbf{Approach}} & \multirow{2}{*}{\textbf{In-the-wild}} & \textbf{FoR} & \textbf{ASV21-LA} & \textbf{ASV21-LA} & \textbf{ASV21-DF} & \textbf{ASV21-DF} \\
    & & &\textbf{Evaluation}  &\textbf{Progress}   &\textbf{Evaluation}   &\textbf{Progress }  &\textbf{Evaluation}  \\
    \midrule
    \textbf{Baseline} & &31.14 &93.04 &11.28 &14.59 &14.21 &20.33\\
    \midrule
    \textbf{0,1\% corresponding} &Supervised &14.43 &68.64 &8.65 &6.44 &3.63 &10.15\\
    \textbf{training data}&Ours &\textbf{8.01} &\textbf{30.12} &\textbf{7.95} &\textbf{5.79} &\textbf{2.69} &\textbf{8.07}\\
    \bottomrule
  \end{tabular}
\end{table*}

\subsection{Problems, Motivation, and Solution}
One of the most concerning problems when dealing with deepfakes in practice, which has not been deeply researched by the community, is maintaining a system that stays up-to-date with the latest deepfake generation techniques. Although our AST has established top performances on unseen training conditions (referenced in Table \ref{tab:itw}\&\ref{tab:fakeavceleb}), its accuracy level still requires significant improvements. This issue may recur as new audio deepfake generation methods emerge. New fake data typically arrive with very few samples, making fine-tuning difficult for deep models. Moreover, this leads to poor model generalization, bias, and, more seriously, the forgetting of previously acquired knowledge. 

\noindent Recently, many applications have witnessed success from LLMs thanks to their Reinforcement learning from human feedback(RLHF) technique. That is, from the model output ranking that feedback from humans, a reward model will be trained, and then from that, that model will be used as a teacher model for the main Language model to learn. Inspired by that, to address this emerging challenge in audio deepfake detection, we propose a plug-in continuous learning method to enhance AST’s performance on new fake data. This semi-supervised plug-in includes a fast-learning detector capable of learning to detect new fake data with a very small number of samples at a reasonable rate. It is followed by the detection of this few-shot learning teacher model on the unsupervised database to collect new labeled samples and update the main AST model after accumulating enough data.

\subsection{Gradient Boosting Approach}
\begin{figure}[h]
  \centering
  \includegraphics[width=\linewidth]{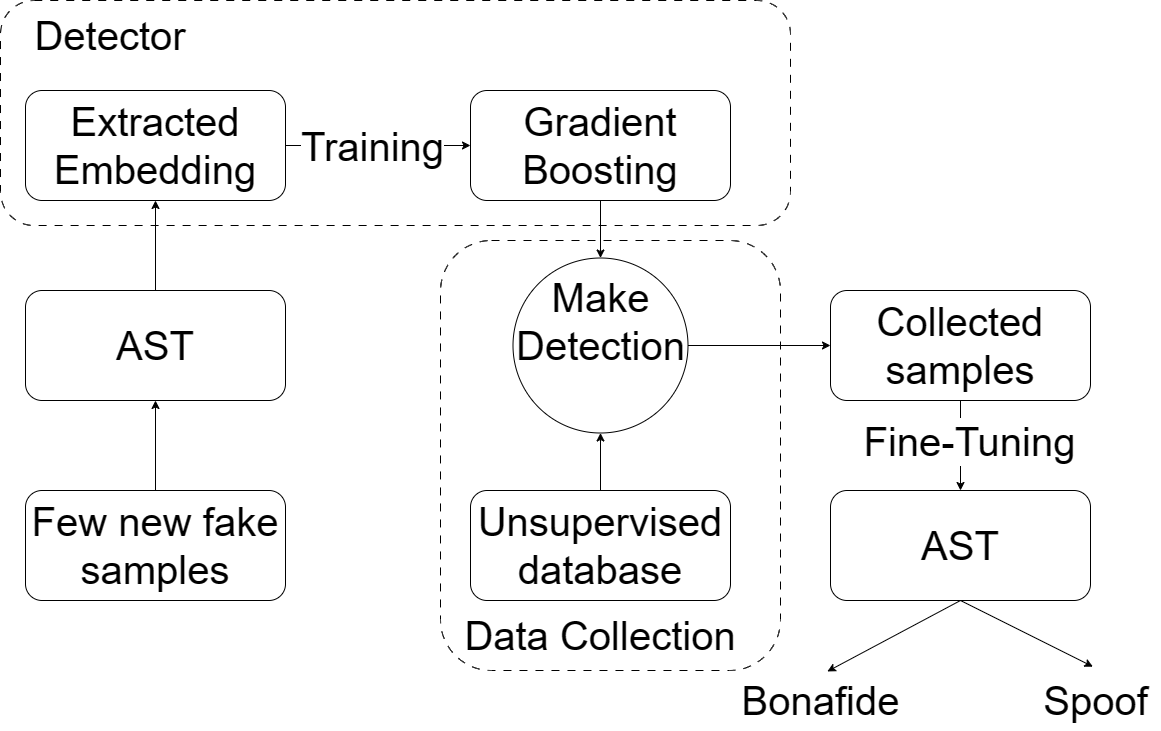}
  \caption{The architecture of the proposed continual learning method.}
  \label{fig:continuous}
\end{figure}

In this paper, a novel plug-in detection based on AST embedding with Gradient Boosting \cite{gradient-boosting} is developed. The block diagram of the processing is illustrated in Fig.\ref{fig:continuous}. 
Initially, we start with the AST model previously trained on 2 million collected samples with augmented data, as described in Section \ref{sec2}. Given only a few samples from a new fake type, the system will utilize a few-shot learning detector based on AST embedding features and an XGBoost classifier \cite{xgboost}. This detector then will be used to collect new fake samples from an unsupervised database and, upon collecting enough (a few hundred, for example), we will then fine-tune the AST model to update the system. For evaluation, we compare the proposed method with a fully supervised approach by setting the initial number of training samples at 0.1\% of the available training dataset and fine-tuning the AST model upon collecting an adequate amount, reaching 5\% of the full training data.

\subsection{Experiments and Results}
\subsubsection{Datasets}
In spite of achieving the best result among all of the current evaluations, our model is still lacking as compared to the SOTA in the in-the-wild dataset. Therefore this remains a good case to try with our continuous learning approach. 

\noindent Specifically, Fake or Real (FoR) \cite{FoR} contains around 111,000 real samples as well as 87,000 bonafide samples. From our findings, the current largest public version has been normalized to have an equal number of samples per class and gender, which makes a total of nearly 70,000 samples. While the synthesis part contains 33 voices from 7 different high-quality TTS systems, real voice samples are both from available data sources and self-collection. The testing part comes from one TTS algorithm that has not been used from training sources. 
\noindent ASVSpoof 2021 is the latest challenge of the ASVSpoof series. The challenge is not focusing on new types of fake but more on dealing with real-life situations of low-quality audio due to transmission or compression. The Logical Access part data source is originally taken from the ASVSpoof 2019 Evaluation part, with the processing of telephony system transmission across 7 different codecs and 4 conditions. Meanwhile, the challenge released an additional DeepFake part which contains data from the 2018 and 2020 voice conversion challenges, focusing on compression conditions by using 9 different compression codecs. 
\noindent The experiments include training XGBoost on 0.1\% of the training set to predict another 5\% of the training set before using these 5\% to fine-tune the model. For in-the-wild and ASVSpoof 2021, since the dataset is not splitting into train and test parts, we conduct the experiment directly on the eval set. In addition, with the ASVSpoof 2021, we are also trying with both the progress along with the evaluation subset. 
\subsubsection{Results analysis}
\noindent In this section, we assess our suggested technique by contrasting its performance in supervised scenarios, which involve direct fine-tuning with as little as 0.1\% of samples, and in unsupervised scenarios, where we apply our method, to showcase the benefits of adopting continual learning. Table \ref{tab:continuous} provides a detailed look at our experimental outcomes on novel datasets using the continuous learning approach we propose. Our findings indicate that incorporating our continuous learning module significantly enhances the ability to detect new, difficult datasets starting from a minimal number of training instances, unlike the traditional fine-tuning approach which proves to be less effective. It's noteworthy to mention that the topic of extensive fine-tuning is beyond the scope of this paper but is intended to be explored in subsequent work.

\section{Conclusions}
\label{sec4}
In this study, we introduce an innovative model for detecting audio deepfakes that leverages the Audio Spectrogram Transformer (AST) architecture, which relies entirely on attention mechanisms without using convolutions. To validate its effectiveness, we amassed a vast dataset of more than 2 million audio deepfake samples, which we further refined with various augmentation methods to enhance its performance in noisy and low-quality audio conditions. Our model set new benchmarks in detecting audio deepfakes across most standard datasets. Additionally, we created an auxiliary model designed to identify newly emerging or difficult types of fake audio. This auxiliary system utilizes AST embeddings from novel datasets, integrating them with an XGBoost classifier to efficiently identify new audio deepfake variations with minimal training samples. It employs a novel technique to create pseudo labels from unlabelled data, allowing the primary model to update itself once sufficient data is collected. Our techniques demonstrated impressive outcomes in tests. Future publications will explore how AST can be applied to real-world scenarios and its impact on continuous model improvement.

\bibliographystyle{IEEEtran}
\bibliography{mybib}

\begin{thebibliography}{10}
\providecommand{\url}[1]{#1}
\csname url@samestyle\endcsname
\providecommand{\newblock}{\relax}
\providecommand{\bibinfo}[2]{#2}
\providecommand{\BIBentrySTDinterwordspacing}{\spaceskip=0pt\relax}
\providecommand{\BIBentryALTinterwordstretchfactor}{4}
\providecommand{\BIBentryALTinterwordspacing}{\spaceskip=\fontdimen2\font plus
\BIBentryALTinterwordstretchfactor\fontdimen3\font minus \fontdimen4\font\relax}
\providecommand{\BIBforeignlanguage}[2]{{%
\expandafter\ifx\csname l@#1\endcsname\relax
\typeout{** WARNING: IEEEtran.bst: No hyphenation pattern has been}%
\typeout{** loaded for the language `#1'. Using the pattern for}%
\typeout{** the default language instead.}%
\else
\language=\csname l@#1\endcsname
\fi
#2}}
\providecommand{\BIBdecl}{\relax}
\BIBdecl

\bibitem{review}
\BIBentryALTinterwordspacing
Z.~Almutairi and H.~Elgibreen, ``A review of modern audio deepfake detection methods: Challenges and future directions,'' \emph{Algorithms}, vol.~15, no.~5, 2022. [Online]. Available: \url{https://www.mdpi.com/1999-4893/15/5/155}
\BIBentrySTDinterwordspacing

\bibitem{spoofing-review}
\BIBentryALTinterwordspacing
Z.~Wu, N.~Evans, T.~Kinnunen, J.~Yamagishi, F.~Alegre, and H.~Li, ``Spoofing and countermeasures for speaker verification: A survey,'' \emph{Speech Communication}, vol.~66, pp. 130--153, 2015. [Online]. Available: \url{https://www.sciencedirect.com/science/article/pii/S0167639314000788}
\BIBentrySTDinterwordspacing

\bibitem{asv17}
\BIBentryALTinterwordspacing
T.~Kinnunen, M.~Sahidullah, H.~Delgado, M.~Todisco, N.~W.~D. Evans, J.~Yamagishi, and K.~Lee, ``The asvspoof 2017 challenge: Assessing the limits of replay spoofing attack detection,'' in \emph{Interspeech 2017, 18th Annual Conference of the International Speech Communication Association, Stockholm, Sweden, August 20-24, 2017}, F.~Lacerda, Ed.\hskip 1em plus 0.5em minus 0.4em\relax {ISCA}, 2017, pp. 2--6. [Online]. Available: \url{http://www.isca-speech.org/archive/Interspeech\_2017/abstracts/1111.html}
\BIBentrySTDinterwordspacing

\bibitem{asv19}
\BIBentryALTinterwordspacing
M.~Todisco, X.~Wang, V.~Vestman, M.~Sahidullah, H.~Delgado, A.~Nautsch, J.~Yamagishi, N.~W.~D. Evans, T.~H. Kinnunen, and K.~A. Lee, ``Asvspoof 2019: Future horizons in spoofed and fake audio detection,'' in \emph{Interspeech 2019, 20th Annual Conference of the International Speech Communication Association, Graz, Austria, 15-19 September 2019}, G.~Kubin and Z.~Kacic, Eds.\hskip 1em plus 0.5em minus 0.4em\relax {ISCA}, 2019, pp. 1008--1012. [Online]. Available: \url{https://doi.org/10.21437/Interspeech.2019-2249}
\BIBentrySTDinterwordspacing

\bibitem{asv21}
\BIBentryALTinterwordspacing
J.~Yamagishi, X.~Wang, M.~Todisco, M.~Sahidullah, J.~Patino, A.~Nautsch, X.~Liu, K.~A. Lee, T.~Kinnunen, N.~Evans, and H.~Delgado, ``Asvspoof 2021: accelerating progress in spoofed and deepfake speech detection,'' 2021. [Online]. Available: \url{https://arxiv.org/abs/2109.00537}
\BIBentrySTDinterwordspacing

\bibitem{wavelet}
A.~Fathan, J.~Alam, and W.~Kang, ``Multiresolution decomposition analysis via wavelet transforms for audio deepfake detection,'' in \emph{Speech and Computer}, S.~R.~M. Prasanna, A.~Karpov, K.~Samudravijaya, and S.~S. Agrawal, Eds.\hskip 1em plus 0.5em minus 0.4em\relax Cham: Springer International Publishing, 2022, pp. 188--200.

\bibitem{ASSERT}
\BIBentryALTinterwordspacing
C.~Lai, N.~Chen, J.~Villalba, and N.~Dehak, ``{ASSERT:} anti-spoofing with squeeze-excitation and residual networks,'' in \emph{Interspeech 2019, 20th Annual Conference of the International Speech Communication Association, Graz, Austria, 15-19 September 2019}, G.~Kubin and Z.~Kacic, Eds.\hskip 1em plus 0.5em minus 0.4em\relax {ISCA}, 2019, pp. 1013--1017. [Online]. Available: \url{https://doi.org/10.21437/Interspeech.2019-1794}
\BIBentrySTDinterwordspacing

\bibitem{resnet34}
P.~Aravind, U.~Nechiyil, N.~Paramparambath \emph{et~al.}, ``Audio spoofing verification using deep convolutional neural networks by transfer learning,'' \emph{arXiv preprint arXiv:2008.03464}, 2020.

\bibitem{wav2vec2}
A.~Baevski, Y.~Zhou, A.~Mohamed, and M.~Auli, ``wav2vec 2.0: A framework for self-supervised learning of speech representations,'' \emph{Advances in neural information processing systems}, vol.~33, pp. 12\,449--12\,460, 2020.

\bibitem{wavlm}
S.~Chen, C.~Wang, Z.~Chen, Y.~Wu, S.~Liu, Z.~Chen, J.~Li, N.~Kanda, T.~Yoshioka, X.~Xiao \emph{et~al.}, ``Wavlm: Large-scale self-supervised pre-training for full stack speech processing,'' \emph{IEEE Journal of Selected Topics in Signal Processing}, vol.~16, no.~6, pp. 1505--1518, 2022.

\bibitem{wavlmdaf}
Y.~Guo, H.~Huang, X.~Chen, H.~Zhao, and Y.~Wang, ``Audio deepfake detection with self-supervised wavlm and multi-fusion attentive classifier,'' \emph{arXiv preprint arXiv:2312.08089}, 2023.

\bibitem{AISG}
\BIBentryALTinterwordspacing
W.~Chen, S.~L.~B. Chua, S.~Winkler, and S.-K. Ng, ``Trusted media challenge dataset and user study,'' in \emph{Proceedings of the 31st {ACM} International Conference on Information {\&} Knowledge Management}.\hskip 1em plus 0.5em minus 0.4em\relax {ACM}, oct 2022. [Online]. Available: \url{https://doi.org/10.1145%2F3511808.3557715}
\BIBentrySTDinterwordspacing

\bibitem{ast}
Y.~Gong, Y.-A. Chung, and J.~Glass, ``{AST: Audio Spectrogram Transformer},'' in \emph{Proc. Interspeech 2021}, 2021, pp. 571--575.

\bibitem{gradient-boosting}
\BIBentryALTinterwordspacing
J.~H. Friedman, ``{Greedy function approximation: A gradient boosting machine.}'' \emph{The Annals of Statistics}, vol.~29, no.~5, pp. 1189 -- 1232, 2001. [Online]. Available: \url{https://doi.org/10.1214/aos/1013203451}
\BIBentrySTDinterwordspacing

\bibitem{self-attention}
\BIBentryALTinterwordspacing
A.~Vaswani, N.~Shazeer, N.~Parmar, J.~Uszkoreit, L.~Jones, A.~N. Gomez, L.~Kaiser, and I.~Polosukhin, ``Attention is all you need,'' 2017. [Online]. Available: \url{https://arxiv.org/pdf/1706.03762.pdf}
\BIBentrySTDinterwordspacing

\bibitem{hubert}
W.-N. Hsu, B.~Bolte, Y.-H.~H. Tsai, K.~Lakhotia, R.~Salakhutdinov, and A.~Mohamed, ``Hubert: Self-supervised speech representation learning by masked prediction of hidden units,'' \emph{IEEE/ACM Transactions on Audio, Speech, and Language Processing}, vol.~29, pp. 3451--3460, 2021.

\bibitem{finetunewav2vec2}
Y.~Wang, A.~Boumadane, and A.~Heba, ``A fine-tuned wav2vec 2.0/hubert benchmark for speech emotion recognition, speaker verification and spoken language understanding,'' \emph{arXiv preprint arXiv:2111.02735}, 2021.

\bibitem{whisper}
A.~Radford, J.~W. Kim, T.~Xu, G.~Brockman, C.~McLeavey, and I.~Sutskever, ``Robust speech recognition via large-scale weak supervision,'' in \emph{International Conference on Machine Learning}.\hskip 1em plus 0.5em minus 0.4em\relax PMLR, 2023, pp. 28\,492--28\,518.

\bibitem{fakeavceleb}
H.~Khalid, S.~Tariq, M.~Kim, and S.~S. Woo, ``Fakeavceleb: A novel audio-video multimodal deepfake dataset,'' \emph{arXiv preprint arXiv:2108.05080}, 2021.

\bibitem{itw}
N.~Müller, P.~Czempin, F.~Diekmann, A.~Froghyar, and K.~Böttinger, ``{Does Audio Deepfake Detection Generalize?}'' in \emph{Proc. Interspeech 2022}, 2022, pp. 2783--2787.

\bibitem{deepfakeSV}
A.~Pianese, D.~Cozzolino, G.~Poggi, and L.~Verdoliva, ``Deepfake audio detection by speaker verification,'' in \emph{2022 IEEE International Workshop on Information Forensics and Security (WIFS)}.\hskip 1em plus 0.5em minus 0.4em\relax IEEE, 2022, pp. 1--6.

\bibitem{fakeavcelebevaluation}
H.~Khalid, M.~Kim, S.~Tariq, and S.~S. Woo, ``Evaluation of an audio-video multimodal deepfake dataset using unimodal and multimodal detectors,'' in \emph{Proceedings of the 1st workshop on synthetic multimedia-audiovisual deepfake generation and detection}, 2021, pp. 7--15.

\bibitem{xgboost}
T.~Chen and C.~Guestrin, ``Xgboost: A scalable tree boosting system,'' in \emph{Proceedings of the 22nd acm sigkdd international conference on knowledge discovery and data mining}, 2016, pp. 785--794.

\bibitem{FoR}
R.~Reimao and V.~Tzerpos, ``For: A dataset for synthetic speech detection,'' 10 2019, pp. 1--10.

\end{thebibliography}

\end{document}